# Sub-100-as timing jitter optical pulse trains from mode-locked Er-fiber lasers

Tae Keun Kim,[1,2] Youjian Song,[1] Kwangyun Jung,[1] Chur Kim,[1] Hyoji Kim,[1] Chang Hee Nam,[2] and Jungwon Kim[1,*]

[1]*KAIST Institute for Optical Science and Technology and School of Mechanical, Aerospace and Systems Engineering, Korea Advanced Institute of Science and Technology (KAIST), Daejeon 305-701, Korea*
[2]*Department of Physics and Coherent X-Ray Research Center, Korea Advanced Institute of Science and Technology (KAIST), Daejeon 305-701, Korea*
*Corresponding author: jungwon.kim@kaist.ac.kr*

We demonstrate sub-100-attosecond timing jitter optical pulse trains generated from free-running, 77.6-MHz repetition-rate, mode-locked Er-fiber lasers. At -0.002($\pm$0.001) ps$^2$ net cavity dispersion, the rms timing jitter is 70 as (224 as) integrated from 10 kHz (1 kHz) to 38.8 MHz offset frequency, when measured by a 24-as-resolution balanced optical cross-correlator. To our knowledge, this result corresponds to the lowest rms timing jitter measured from any mode-locked fiber lasers so far. The measured result also agrees fairly well with the Namiki-Haus analytic model of quantum-limited timing jitter in stretched-pulse fiber lasers.

Femtosecond mode-locked fiber lasers are finding more and more applications, ranging from materials processing to precision measurements, owing to their excellent characteristics such as high gain, low noise, good thermal property, compactness, and easiness in building and operation. In particular, the excellent noise properties of mode-locked fiber lasers enable various time-frequency applications. In the frequency domain, the extremely low optical phase noise of mode-locked fiber lasers has enabled sub-mHz linewidth level frequency combs [1]. In the time domain, the fluctuation of pulse positions from perfectly periodic positions, i.e., timing jitter, has been expected to be extremely low as well. The quantum-limited timing jitter of free-running mode-locked lasers has been expected to be well below a femtosecond [2-5]. Due to their inherently low level, only very recently, accurate measurement of timing jitter up to the Nyquist frequency has been possible by using the balanced optical cross-correlation (BOC) method [6-8]. The use of attosecond-resolution BOC method has enabled the demonstration of 20-as-level jitter from Ti:sapphire solid-state lasers at 800 nm [9] and 200-as-level jitter from Yb-fiber lasers at 1 μm [10].

The availability of mode-locked Er-fiber lasers at telecomm wavelength (1.5-1.6 μm) with attosecond timing jitter enables wide range of new applications in optical communications and photonic signal processing such as analog-to-digital conversion and long-distance clock distribution via fiber optic links. So far the best timing jitter performance demonstrated for free-running Er-fiber lasers is 2.6 fs (integrated from 10 kHz to 40 MHz) when the laser is operated at the stretched-pulse condition with positive net cavity dispersion [7]. The optimization of timing jitter in Er-fiber lasers toward the attosecond regime has not yet been shown.

In this Letter we demonstrate sub-100-as timing jitter optical pulse trains from 78-MHz free-running mode-locked Er-fiber lasers. The high-frequency timing jitter is scaled down by a factor of ~40 from few-fs [7] to sub-100-as level by a simple cavity dispersion control. At -0.002 ($\pm$0.001) ps$^2$ net cavity dispersion at 1582 nm center wavelength, the rms timing jitter is measured to be 70 as (224 as) integrated from 10 kHz (1 kHz) to 38.8 MHz offset frequency. To our knowledge, this result corresponds to the lowest rms timing jitter measured up to the Nyquist frequency from any mode-locked fiber lasers. Using directly measured laser parameters, we further find that the measured result agrees fairly well (within a few dB difference) with the prediction of the Namiki-Haus analytic noise model [3] for quantum-limited timing jitter in stretched-pulse fiber lasers.

For timing jitter characterization and optimization, two almost identical, ~78 MHz repetition rate, nonlinear polarization evolution (NPE)-based Er-fiber lasers were built (Fig. 1(a)). The net cavity dispersion is set to the close-to-zero range (-0.006 ps$^2$ to +0.002 ps$^2$ in this work) by balancing the positive dispersion of Er-gain fiber (Liekki Er80-4/125) and the negative dispersion of standard single-mode fiber (SMF-28). For the jitter characterization, we put a special effort to make two independent lasers running with almost identical laser parameters. The only difference between the two lasers is the geometry in free-space section: a sigma-cavity with a piezoelectric transducer (PZT)-mounted mirror is employed in one laser (Laser 2 in Fig. 1(a)) to enable repetition-rate locking between the two lasers. The output optical spectra and interferometric autocorrelation (IAC) traces of the two lasers at -0.002($\pm$0.001) ps$^2$ net cavity dispersion are shown in Fig. 1(b). Both lasers have center wavelength of 1582 nm and full-width half-maximum (FWHM) output pulsewidth of ~60 fs.

The timing jitter is characterized by a 24-as resolution PPKTP-based BOC in a similar way as shown in refs. [6] and [7]. In order to confine the jitter measurement in the linear detection range of BOC output (inset graph in Fig. 1(a)), the repetition rate of two lasers is locked by a low-bandwidth phase-locked loop (PLL) using the PZT-mounted mirror in Laser 2 (Fig. 1(a)). The laser locking is achieved by optical locking using the BOC output or by electronic locking using the photo-detected microwave signals (the 13th harmonic of repetition rate at ~1 GHz). Although the optical locking provides tighter long-term stable locking, it requires a sufficient locking bandwidth (at least a few kHz) to acquire and maintain the locking because of a large resonant peak near locking bandwidth. This resonant peak was inevitable to fully suppress strong acoustic noise peaks in the 100 – 300 Hz range, which are induced by instrument fan noise and other vibration noise sources. Combined with the limited locking range of the BOC (~80 fs peak-to-peak), stable optical locking was obtained only when the locking bandwidth is at least a few kHz, which results in a valid jitter measurement from ~10 kHz offset frequency [7,8]. In order to measure timing jitter in the lower offset frequency range (below 10 kHz), we also use the electronic locking with ~100 Hz locking bandwidth and ~3 fs (rms) residual jitter, which enables the jitter measurement in the linear detection range of the BOC. Using these locking techniques, the measurement of timing jitter spectrum over 5 decades of offset frequency from 300 Hz to ~39 MHz (Nyquist frequency) is possible. An RF and an FFT analyzers are used to measure the timing jitter spectral density of the BOC output in the 10 kHz–39 MHz and 300 Hz–10 kHz ranges, respectively. To get the timing jitter of a single laser, the measured jitter spectral density is divided by two because two lasers are almost identical and uncorrelated [7].

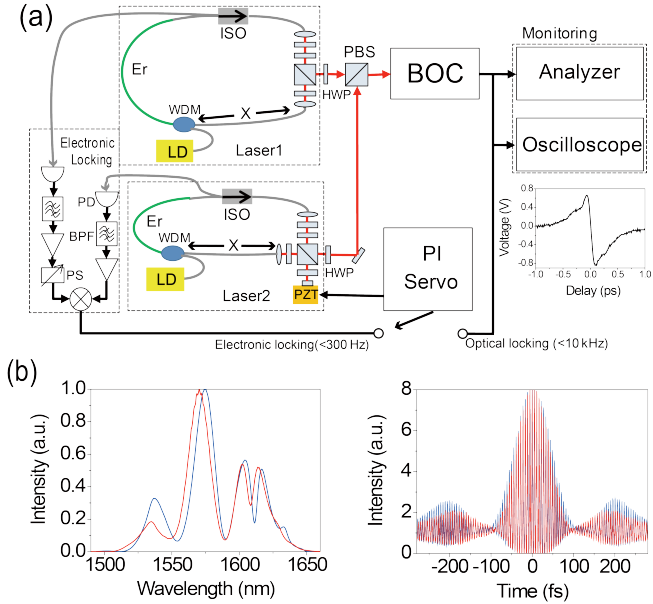

Fig. 1. (color online) (a) Experimental setup for timing jitter characterization and optimization of mode-locked Er-fiber lasers. BOC, balanced optical cross-correlator; BPF, 1-GHz bandpass filter; Er, Er-gain fiber; ISO, isolator; PD, photodetector; PS, phase shifter. The net cavity dispersion of each laser is controlled by changing the length of single-mode fiber (labelled as X). (b) Optical spectra and interferometric autocorrelation of lasers (red: Laser 1, blue: Laser 2).

In order to find the lowest jitter condition, dispersion control is performed by adding or cutting SMF-28 fiber inside the laser cavity (Fig. 1(a) 'X' section). The net-cavity dispersion is tuned in the near-zero and slightly negative dispersion regime ranging from -0.006 $ps^2$ to +0.002 $ps^2$ with ~0.001 $ps^2$ interval at 1582 nm center wavelength by adding or cutting SMF-28 fiber by ~5 cm interval. The net-cavity dispersion is measured by an in-situ dispersion measurement method as used in refs. [10] and [11]. Within a rather narrow range of dispersion from -0.006 $ps^2$ to +0.002 $ps^2$, the measured timing jitter spectral density level changed by more than 20 dB, similar to the recent result for Yb-fiber lasers [10]. Such dramatic change in jitter is mainly due to the change in pulsewidth and pulse chirp for different cavity dispersion condition. The contribution of slight change in repetition rate by adding or cutting fiber segments to the laser jitter is negligible. For the Er-fiber lasers used in this experiment, the lowest timing jitter is achieved when the intra-cavity dispersion is set to -0.002(±0.001) $ps^2$ at 1582 nm center wavelength.

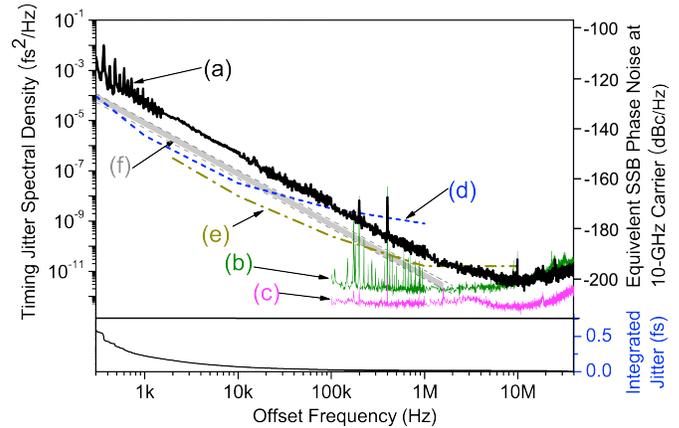

Fig. 2. (color online) (a) The best timing jitter spectral density result of mode-locked Er-fiber lasers. The integrated timing jitter is 70 as (224 as) when integrated from 10 kHz (1 kHz) to 38.8 MHz offset frequency. (b) BOC photodetector noise floor. (c) Projected RIN-coupled timing jitter. (d) The equivalanet timing jitter spectral density of SLCO [12] for comparison. (e) The best timing jitter spectral density of mode-locked Ti:sapphire lasers [9] for comparison. (f) Predicted timing jitter range from Namiki-Haus analytic model based on measured laser parameters.

The measured lowest timing jitter spectral density is shown in Fig. 2 (curve (a)). The jitter spectral density shows a $1/f^2$ slope from 300 Hz to 2 MHz offset frequency, which is a characteristic of random walk nature of the timing jitter in a free-running laser. In the high offset frequency above 8 MHz, the jitter measurement is limited by the BOC photodetector noise (at ~3×$10^{-12}$ $fs^2$/Hz level, curve (b) in Fig. 2). If a higher-resolution measurement is possible, the high-frequency jitter spectrum will be limited by the relative intensity noise (RIN)-coupled timing jitter from Kramers-Kronig relationship [4,10]. From the RIN measurement of the laser used, we find that the projected RIN-coupled jitter contribution (curve (c) in Fig. 2) is slightly lower than the detector noise (curve (b) in Fig. 2). The integrated rms timing jitter from 10 kHz (1 kHz) to

38.8 MHz offset frequency is 70 as (224 as). To our knowledge, this is the lowest timing jitter ever achieved for free-running mode-locked fiber lasers so far. For comparison, we also plot the equivalent timing jitter of commercial microwave signal generators with the lowest phase noise (sapphire loaded cavity oscillator (SLCO) [12], curve (d) in Fig.2) and the lowest timing jitter result of 80-MHz Ti:sapphire lasers [9] (curve (e) in Fig.2).

We further compare the measured jitter spectrum with the Namiki-Haus analytic model of quantum-limited timing jitter in stretched-pulse fiber lasers [3]. To compute the theoretical jitter accurately, we measured most of laser parameters used in the model. The measured and typically-used laser parameters are summarized in Table 1. A Gaussian-shaped pulse with the reference pulsewidth $\tau_0$ of 36 fs ($60~\text{fs}/2\sqrt{\ln 2}$) is determined by the PICASO algorithm [13] using the measured optical spectrum and IAC trace. Intra-cavity pulse energy $w$ (1.6 nJ) is obtained by measuring the repetition rate and the average power from a 5 % output port located right after the Er-gain fiber. The saturated amplitude gain $g$ (0.39) is evaluated by compensating the total cavity loss (55 %) per round-trip in the steady-state. To determine excess noise factor of gain medium $\Theta$, we built an Er-doped fiber amplifier (EDFA) with Er-fiber length and pump power level identical to those in the fiber laser. The excess noise factor $\Theta$ is determined by measuring the amplified spontaneous emission (ASE) spectral density at a particular gain level in the EDFA, similar to the method shown in [14], which results in 14 (±0.2). The proportionality factor α, which is the ratio of self-amplitude modulation (SAM) to self-phase modulation (SPM), typically ranges from 0.1 to 0.3 in NPE-based fiber lasers [3], and we use this range of α in the model. The chirp parameter β is accordingly calculated from [3]

$$\beta = \tan\left\{\frac{1}{2}\left[\arg(\alpha - j) - \arg(\frac{g}{\Omega_g} + jD)\right]\right\},$$

which results in β ranging from -0.52 to -0.12. Using these parameters, predicted jitter spectra of the Namiki-Haus model are shown by gray region (f) in Fig. 2. The upper and lower bound corresponds to the parameter combinations of ($D$, α, β, $\Theta$) = (-0.003 ps², 0.1, -0.22, 14.2) and (-0.001 ps², 0.3, -0.40, 13.8), respectively. The measured jitter spectrum (curve (a) in Fig. 2) agrees fairly well with the prediction of analytic model within 5–8 dB difference. The discrepancy of a few dB between the measurement and the model might be due to higher order effects such as third-order dispersion and nonlinear chirp present in the fiber laser.

In summary, we have scaled down the timing jitter of mode-locked Er-fiber lasers toward the sub-100-as regime by intra-cavity dispersion control. At -0.002(±0.001) ps² dispersion condition the integrated rms timing jitter from 10 kHz (1 kHz) to 38.8 MHz is measured to be 70 as (224 as), which is a factor of ~40 improvement from the previously reported result [7]. As shown in Fig. 2, the measured timing jitter spectrum of Er-fiber lasers is comparable to those of the best microwave sources [12] and Ti:sapphire lasers [9]. Although Ti:sapphire lasers can provide lower jitter (~20 as) due to their much shorter pulsewidth (<10 fs) and less chirp, the Er-fiber lasers with ~70 as jitter performance demonstrated in this work are expected to be attractive photonic signal sources due to their lower cost, simpler design and implementation, and the compatibility with 1550-nm telecomm devices and fibers. We further show that the Namiki-Haus analytic model based on measured laser parameters can predict the jitter spectrum fairly well within a few dB difference, which suggests that one can effectively use an existing analytic model for optimizing the fiber laser design for the targeted jitter performances.

Table 1. Measured and typically-used laser parameters for Namiki-Haus model

| Measured laser parameters | |
| --- | --- |
| Reference pulsewidth, $\tau_0$ (fs) | 36 |
| Center wavelength (nm) | 1582 |
| Intra-cavity pulse energy, $w$ (nJ) | 1.6 |
| Saturated amplitude gain, $g$ | 0.39 |
| Excess noise factor, $\Theta$ | 14 (±0.2) |
| Net intra-cavity dispersion, $D$ (ps²) | -0.002 (±0.001) |
| **Typically-used laser parameters** | |
| HWHM gain bandwidth, $\Omega_g$ (rad/s) | $1.51 \times 10^{13}$ [7] |
| SAM-SPM ratio, $\alpha$ | 0.1 ~ 0.3 [3] |
| Chirp parameter, $\beta$ | -0.52 ~ -0.12 [3] |

We thank Jonathan A. Cox for fruitful discussions on Er-fiber lasers and their locking. This research was supported by the National Research Foundation of Korea (NRF) funded by the Ministry of Education, Science and Technology (2010-0003974).